\documentclass[12pt,preprint]{aastex}

\usepackage{color}
\usepackage{url}
\usepackage{graphicx}
\graphicspath{{figures/}}

\usepackage{listings}
\definecolor{lbcolor}{rgb}{0.9,0.9,0.9}
\newcommand{\foreign}[1]{{\it #1}}

\newcommand{\apriori}{\foreign{a priori}}
\newcommand{\adhoc}{\foreign{ad hoc}}

\newcommand{\Fig}[1]{Figure~\ref{fig:#1}}
\newcommand{\fig}[1]{\Fig{#1}}
\newcommand{\figlabel}[1]{\label{fig:#1}}
\newcommand{\Eq}[1]{Equation~(\ref{eq:#1})}
\newcommand{\eq}[1]{\Eq{#1}}
\newcommand{\eqlabel}[1]{\label{eq:#1}}
\newcommand{\Sect}[1]{Section~\ref{sect:#1}}
\newcommand{\sect}[1]{\Sect{#1}}
\newcommand{\App}[1]{Appendix~\ref{sect:#1}}
\newcommand{\app}[1]{\App{#1}}
\newcommand{\sectlabel}[1]{\label{sect:#1}}

\begin{document}

\title{Periodograms for Multiband Astronomical Time Series}

\newcommand{\escience}{1}
\newcommand{\uwastro}{2}
\author{Jacob T. VanderPlas\altaffilmark{\escience}}
\author{{\v Z}eljko Ivezi{\'c}\altaffilmark{\uwastro}}
\altaffiltext{\escience}{eScience Institute, University of Washington}
\altaffiltext{\uwastro}{Department of Astronomy, University of Washington}

\begin{abstract}
This paper introduces the {\it multiband periodogram}, a general extension of the well-known 
Lomb-Scargle approach for detecting periodic signals in time-domain data. In addition to 
advantages of the Lomb-Scargle method such as treatment of non-uniform sampling and
heteroscedastic errors, the multiband periodogram significantly improves period finding for randomly 
sampled multiband light curves (e.g., Pan-STARRS, DES and LSST). The light curves in 
each band are modeled as arbitrary truncated Fourier series, with the period and phase 
shared across all bands. The key aspect is the use of Tikhonov regularization which drives
most of the variability into the so-called base model common to all bands, while 
fits for individual bands describe residuals relative to the base model and typically require
lower-order Fourier series. This decrease in the effective model complexity is the
main reason for improved performance. We use simulated light curves 
and randomly subsampled SDSS Stripe 82 data to demonstrate the superiority of this 
method compared to other methods from the literature, and find that this method will be able to
efficiently determine the correct period in the majority of LSST's bright RR Lyrae stars with
as little as six months of LSST data.
A Python implementation of this method, along with code to fully reproduce the results
reported here, is available on GitHub.
\end{abstract}


\section{Introduction}
\sectlabel{introduction}

Many types of variable stars show periodic flux variability \citep{EM2008}. Periodic variable stars are important 
both for testing models of stellar evolution and for using such stars as distance indicators (e.g., Cepheids 
and RR Lyrae stars). One of the first and main goals of the analysis is to detect variability and to estimate the 
period and its uncertainty. A number of parametric and non-parametric methods have been proposed to 
estimate the period of an astronomical time series \citep[e.g.,][and references therein]{Graham13}.

The most popular non-parametric method is the phase dispersion minimization (PDM) introduced by \cite{PDM1978}. 
Dispersion per bin is computed for binned phased light curves evaluated for a grid of trial periods. The best
period minimizes the dispersion per bin.  A similar and related non-parametric method that has been recently 
gaining popularity is the Supersmoother routine \citep{Reimann94}. It uses a running mean or running linear 
regression on the data to fit the observations as a function of phase to a range of periods. The best period 
minimizes a figure-of-merit, adopted as weighted sum of absolute residuals around the running mean. 
Neither the Supersmoother algorithm nor the PDM method require \apriori{} knowledge of the light curve shape. 
We note that neither method produces posterior probability distribution for the period but rather a single point 
estimate. 

The most popular parametric method is the Lomb-Scargle periodogram, which is discussed in detail in \sect{brief_overview}.
The Lomb-Scargle periodogram is related to the $\chi^2$ for a least-square fit of a single sinusoid to data
and can treat non-uniformly sampled time series with heteroscedastic measurement uncertainties. 
The underlying model of the Lomb–Scargle periodogram is nonlinear in frequency and basis functions at different
frequencies are not orthogonal. As a result, the periodogram has many local maxima and thus in practice the global 
maximum of the periodogram is found by grid search \citep[for details see, e.g.][]{ICVG2014}.
A more general parametric method based on the use of continuous-time autoregressive moving average (CARMA) model
was recently introduced by \citet{Kelly14}. CARMA models can also treat non-uniformly sampled time series with 
heteroscedastic measurement uncertainties, and can handle complex variability patterns. 

A weakness of the above methods is that they require homogeneous measurements -- for astronomy data, this means 
that successive measurements must be taken through a single photometric bandpass (filter). This has not been a major
problem for past surveys because measurements are generally taken through a single photometric filter 
\citep [e.g. LINEAR,][]{LINEAR1}, or nearly-simultaneously in all bands at each observation \citep [e.g. SDSS,][]{Sesar2010}.
For the case of simultaneously taken multiband measurements, \cite{Suveges12} utilized the principal component
method to optimally extract the best period. Their method is essentially a multiband generalization of the well-known
two-band Welch-Stetson variability index \citep{Stetson1996}. Unfortunately, when data in each band are taken at
different times, such an approach in not applicable. In such cases, past studies have generally relied 
on \adhoc{} methods such as a majority vote among multiple single-band estimates of the 
periodogram \citep[e.g.,][]{Oluseyi12}. 

For surveys that obtain multiband data one band at a time, such as Pan-STARRS \citep{Kaiser2010} and DES \citep{Flaugher08},
and for future multicolor surveys such as LSST \citep{Ivezic08LSST}, this \adhoc{} approach is not optimal. In order to take 
advantage of the full information content in available data, it would be desirable to have a single estimate of the periodogram 
which accounts for all observed data in a manner which does not depend on the underlying spectrum of the object. 
We propose such a method in this paper. 

The proposed method is essentially a generalization of the Lomb-Scargle method to 
multiband case. The light curves in  each band are modeled as arbitrary truncated Fourier series, 
with the period, and optionally the phase, shared across all bands. The key aspect enabling this approach is the use of Tikhonov regularization 
(discussed in detail in \sect{regularization}) which drives most of the variability into the so-called {\it base 
model} common to all bands, while fits for individual bands describe residuals relative to the base model 
and typically require lower-order Fourier series. This decrease in effective model complexity is the
main reason for improved performance. 

The remainder of the paper is organized as follows:
in \sect{brief_overview} we provide a brief review of least-squares periodic fitting, and in \sect{matrix_formalism} derive the matrix-based formalism for single-band periodic analysis used through the rest of this work.
\sect{extending_periodogram} introduces several extensions and generalizations to the single-band model that the matrix formalism makes possible, including floating mean models, truncated Fourier models, and regularized models.
In \sect{multiband}, we use the ideas behind these extensions to motivate the {\it multiband periodogram}, and show some examples of its use on simulated data.
In \sect{stripe82} we apply this method to measurements of 483 RR Lyrae stars first explored by \citet[][hereafter S10]{Sesar2010}, and in \sect{LSST} explore the performance of the method for simulated observations from the LSST survey.
We conclude in \sect{discussion}.

\section{Brief Overview of Periodic Analysis}
\sectlabel{brief_overview}

The detection and quantification of periodicity in time-varying signals is an important area of data analysis within modern time-domain astronomical surveys.
For evenly-spaced data, the {\it periodogram}, a term coined by \citet{Schuster98}, gives a quantitative measure of the periodicity of data as a function of the angular frequency $\omega$. For data $\{y_k\}_{k=1}^N$ measured at equal intervals $t_k = t_0 + k\Delta t$, Schuster's periodogram, which measures the spectral power as a function of the angular frequency, is given by
\begin{equation}
  \eqlabel{Schuster}
  C(\omega) = \frac{1}{N}\left| \sum_{k=1}^N y_k e^{i\omega t_k} \right|^2,
\end{equation}
and can be computed very efficiently using the Fast Fourier Transform.

Because astronomical observing cadences are rarely so uniform, many have looked at extending the ideas behind the periodogram to work with unevenly-sampled data. Most famously, \citet{Lomb76} and \citet{Scargle82} extended earlier work to define the {\it normalized periodogram}:
\begin{equation}
  \eqlabel{LombScargle}
  P_N(\omega) = \frac{1}{2\,V_y}\left[
    \frac{\left[\sum_k(y_k - \bar{y})\cos\omega(t_k - \tau)\right]^2}
    {\sum_k \cos^2\omega(t_k - \tau)}
    +
    \frac{\left[\sum_k(y_k - \bar{y})\sin\omega(t_k - \tau)\right]^2}
    {\sum_k \sin^2\omega(t_k - \tau)}
\right],
\end{equation}
where $\bar{y}$ is the mean and $V_y$ is the variance of the data $\{y_k\}$, and $\tau$ is the time-offset which makes $P_N(\omega)$ independent of a translation in $t$ \citep[see][for an in-depth discussion]{NumRec}. \citet{Lomb76} showed that this time-offset has a deeper effect: namely, it makes $P_N$ identical to the estimate of harmonic content given a least-squares fit to a single-component sinusoidal model,
\begin{equation}
  \eqlabel{SingleModel}
  d(t) = A\sin(\omega t + \phi).
\end{equation}
This long-recognized connection between spectral power and least squares fitting methods was solidified by \citet{Jaynes87}, who demonstrated that the normalized periodogram of Lomb and Scargle is a sufficient statistic for inferences about a stationary-frequency signal in the presence of Gaussian noise. Building on this result, \citet{Bretthorst88} explored the extension of these methods to more complicated models with multiple frequency terms, non-stationary frequencies, and other more sophisticated models within a Bayesian framework.

While the important features of least squares frequency estimation via Lomb-Scargle periodograms have been discussed elsewhere, we will present a brief introduction to the subject in the following section.
The matrix-based formalism we develop here will make clear how the method can be extended to more sophisticated models, including the multiband periodogram proposed in this work.

\section{Standard Least Squares Spectral Fitting}
\sectlabel{matrix_formalism}

In this section we present a brief quantitative introduction to the least squares fitting formulation of the normalized periodogram of \eq{LombScargle}. We denote $N$ observed data points as
\begin{equation}
  D = \{t_k, y_k, \sigma_k\}_{k=1}^N
\end{equation}
where $t_k$ is the time of observation, $y_k$ is the observed value (typically a magnitude), and $\sigma_k$ describes the Gaussian errors on each value. Without loss of generality we will assume that the data $y_k$ are centered such that the measurements within each band satisfy
\begin{equation}
  \eqlabel{ycentered}
  \frac{\sum_k w_ky_k}{\sum_k w_k} = 0
\end{equation}
where the weights are $w_k = \sigma_k^{-2}$.

\subsection{Stationary Sinusoid Model}

The normalized periodogram of \eq{LombScargle} can be derived from the normalized $\chi^2$ of the best-fit single-term stationary sinusoidal model given in \eq{SingleModel}. To make the problem linear, we can re-express the model in terms of the parameter vector $\theta = [A\cos\phi, A\sin\phi]$ so that our model is
\begin{equation}
  \eqlabel{simplemodel}
  y(t|\omega,\theta) = \theta_1\sin(\omega t) + \theta_2\cos(\omega t).
\end{equation}
For a given $\omega$, the maximum likelihood estimate of the parameters $\theta$ can be found by minimizing the $\chi^2$ of the model, which is given by
\begin{equation}
  \chi^2(\omega) = \sum_k \frac{[y_k - y(t_k|\omega,\theta)]^2}{\sigma_k^2}.
\end{equation}
For the single-term Fourier model, it can be shown \citep[see, e.g.][]{ICVG2014} that
\begin{equation}
  \eqlabel{chi2PN}
  \chi_{min}^2(\omega) = \chi^2_0[1 - P_N(\omega)]
\end{equation}
where $P_N(\omega)$ is the normalized periodogram given in \eq{LombScargle} and $\chi^2_0$ is the reference $\chi^2$ for a constant model, which due to the assumption in \eq{ycentered} is simply $\chi^2_0 = \sum_k (y_k/\sigma_k)^2$.

\subsection{Matrix Formalism}
The expressions related to the stationary sinusoid model can be expressed more compactly by defining the following matrices:
\begin{equation}
X_\omega = \left[
\begin{array}{cc}
\sin(\omega t_1) & \cos(\omega t_1)\\
\sin(\omega t_2) & \cos(\omega t_2)\\
\vdots & \vdots \\
\sin(\omega t_N) & \cos(\omega t_N)\\
\end{array}
\right];~~
y = \left[
\begin{array}{c}
y_1 \\
y_2\\
\vdots \\
y_N\\
\end{array}
\right];~~
\Sigma = \left[
\begin{array}{cccc}
\sigma_1^2 & 0 &  \cdots & 0\\
0 & \sigma_2^2 &  \cdots & 0\\
\vdots & \vdots &  \ddots & \vdots\\
0 & 0 &  \cdots & \sigma_N^2
\end{array}
\right]
\end{equation}
With these definitions, the model in \eq{simplemodel} can be expressed as a simple linear product, $y(t|\omega,\theta) = X_\omega\theta$, and the model and reference $\chi^2$ can be written

\begin{eqnarray}
  \chi^2(\omega) &=& (y - X_\omega\theta)^T\Sigma^{-1}(y - X_\omega\theta)\\
  \chi^2_0 &=& y^T \Sigma^{-1} y
\end{eqnarray}
The expression for the normalized periodogram can be computed by finding via standard methods the value of $\theta$ which minimizes $\chi^2(\omega)$, and plugging the result into \eq{chi2PN}. This yields
\begin{equation}
  \eqlabel{LombScargle2}
  P_N(\omega) = \frac{y^T\Sigma^{-1}X_\omega~[X_\omega^T\Sigma^{-1}X_\omega]^{-1}~X_\omega^T\Sigma^{-1}y}{y^T\Sigma^{-1}y}.
\end{equation}
We note that this expression is equivalent to \eq{LombScargle} in the homoscedastic case with $\Sigma \propto V_y I$.

\subsection{Simple Single-band Period Finding}
\sectlabel{simple_period}

\begin{figure}
  \centering
  \includegraphics[width=\textwidth]{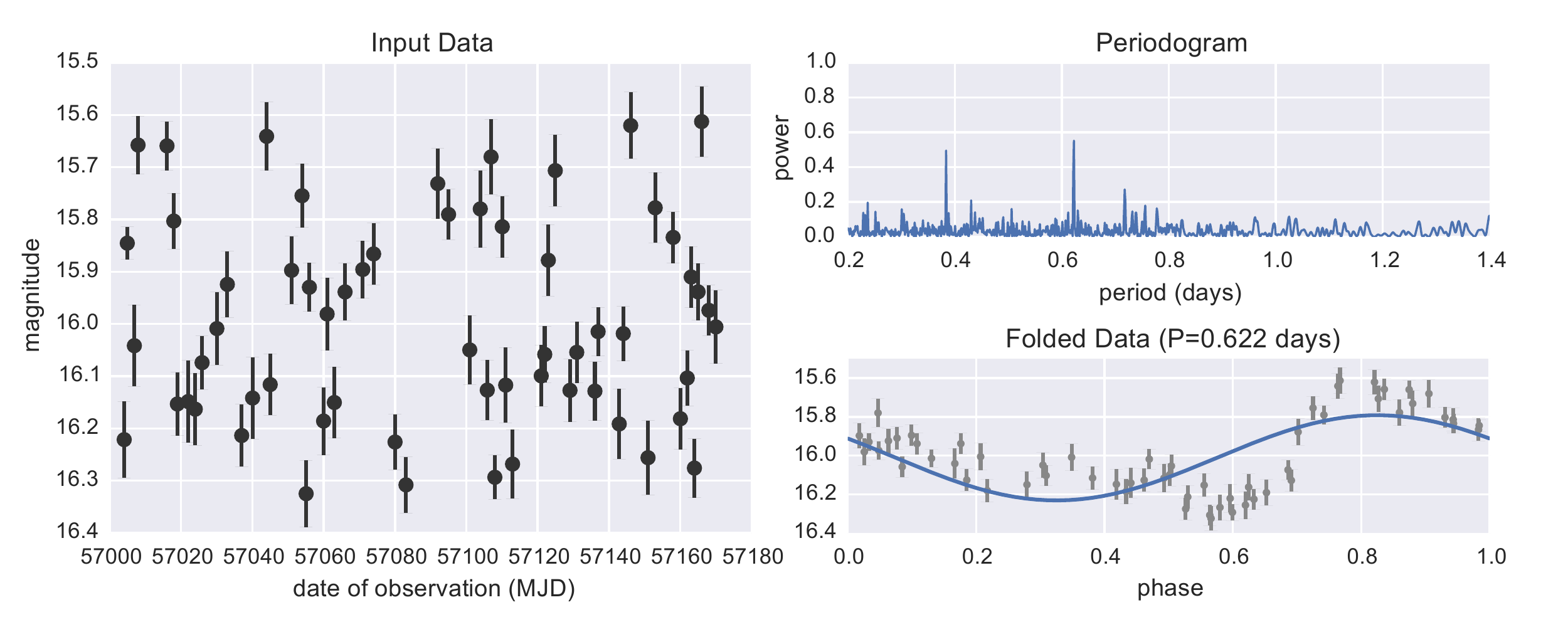}
  \caption{
    An illustration of the basic periodogram and its relationship to the single-term sinusoid model. The left panel shows the input data, while the right panels show the fit derived from the data. The upper-right panel shows the periodogram with a clear peak at the true period of 0.622 days, and the bottom-right panel shows the data as a function of the phase associated with this period. Note in the periodogram the presence of the typical aliasing effect, with power located at beat frequencies between the true period and the 1-day observing cadence (see \sect{simple_period} for further discussion).
  }
  \figlabel{basic_example}
\end{figure}

As an example of the standard periodogram in action, we perform a simple single-band harmonic analysis of simulated $r$-band observations of an RR Lyrae light curve, based on empirical templates derived in S10 (\fig{basic_example}). The observations are of a star with a period of 0.622 days, and take place on 60 random nights over a 6-month period, as seen in the left panel.

The upper-right panel shows the normalized periodogram for this source as a function of period. While the power does peak at the true period of 0.622 days, an aliasing effect is readily apparent near $P=0.38$. This additional peak is due to beat frequency between the true period $P$ and the observing cadence of $\sim 1$ day. This beat frequency is the first in a large sequence: for nightly observations, we'd expect to find excess power at periods $P_n = P / (1 + nP)$ days, for any integer $n$. The strong alias in \fig{basic_example} corresponds to the $n=1$ beat period $P_n=0.383$. Though it is possible to carefully correct for such aliasing by iteratively removing contributions from the estimated window function \citep[e.g.][]{Roberts87}, we'll ignore this detail in the current work.

The lower-right panel of \fig{basic_example} shows the maximum likelihood interpretation of this periodogram: it is a measure of the normalized $\chi^2$ for a single-term sinusoidal model. Here we visualize the data from the left panel, but folded as a function of phase, and overplotted with the best-fit single-term model. This visualization makes it apparent that the single-term model is highly biased: RR Lyrae light curves are, in general, much more complicated than a simple sinusoid. Nevertheless, the simplistic sinusoidal model is able to recover the correct frequency to a high degree of accuracy (roughly related to the width of the peak) and significance (roughly related to the height of the peak). For a more complete introduction to and discussion of the single-term normalized periodogram, refer to, e.g. \citet{Bretthorst88} or \citet{ICVG2014}.

\section{Extending the Periodogram}
\sectlabel{extending_periodogram}
We have shown two forms of the classic normalized periodogram: \eq{LombScargle} and \eq{LombScargle2}. Though the two expressions are equivalent, they differ in their utility. Because the expression in \eq{LombScargle} avoids the explicit construction of a matrix, it can be computed very efficiently. Furthermore, through clever use of the Fast Fourier Transform, expressions of the form of \eq{LombScargle} can be evaluated exactly for $N$ frequencies in $\mathcal{O}[\log{N}]$ time \citep{Press89}.

The matrix-based formulation of \eq{LombScargle2}, though slower than the Fourier-derived formulation, is a more general expression and allows several advantages:
\begin{enumerate}
  \item It is trivially extended to heteroscedastic and/or correlated measurement noise in the data $y_k$ through appropriate modification of the {\it noise covariance matrix} $\Sigma$.
  \item It is trivially extended to more sophisticated linear models by appropriately modifying the {\it design matrix} $X_\omega$.
  \item It is trivially extended to include Tikhonov/L2-regularization terms (see \sect{regularization} for more details)  by adding an appropriate diagonal term to the {\it normal matrix} $X_\omega^T\Sigma^{-1}X_\omega$.
\end{enumerate}
In the remainder of this section, we will explore a few of these modifications and how they affect the periodogram and resulting model fits.

\subsection{Stationary Sinusoid with Floating Mean}
\sectlabel{floating_mean}

\begin{figure}
  \centering
  \includegraphics[width=\textwidth]{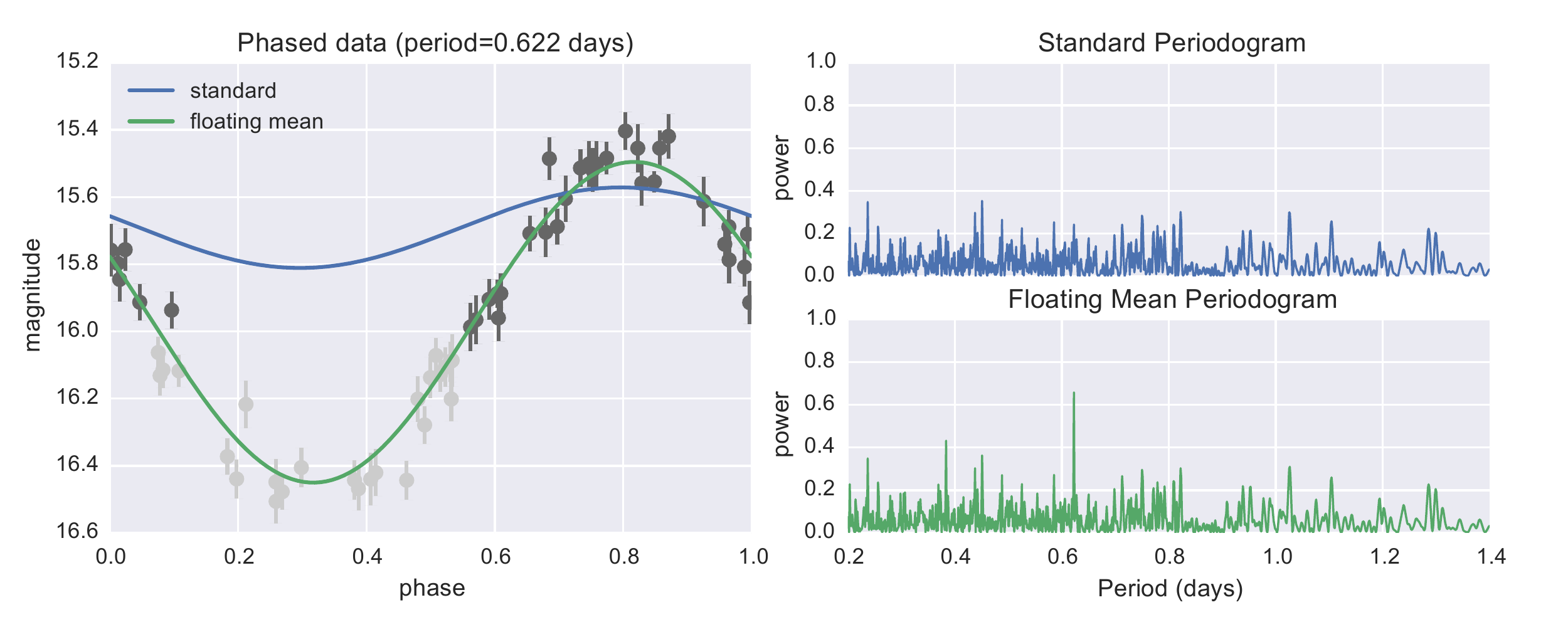}
  \caption{
    An illustration of the effect of the floating mean model for censored data.
    The data consist of 80 observations drawn from a sinusoidal model. To mimic a potentially damaging selection effect, all observations with magnitude fainter than 16 are removed (indicated by the light-gray points). The standard and floating-mean periodograms are computed from the remaining data; these fits are shown over the data in the left panel. Because of this biased observing pattern, the mean of the observed data is a biased estimator of the true mean. The standard fixed-mean model in this case fails to recover the true period of 0.622 days, while the floating mean model still finds the correct period.
  }
  \figlabel{floating_mean}
\end{figure}

As an example of one of these generalizations, we'll consider what \citet{Zechmeister09} call the ``generalized Lomb-Scargle method'', and which we'll call the {\it floating-mean periodogram}. This method adjusts the classic normalized periodogram by fitting the mean of the model alongside the amplitudes:
\begin{equation}
  y(t~|~\omega, \theta) = \theta_0 + \theta_1\sin\omega t + \theta_2\cos\omega t
\end{equation}
This model can be more accurate than the standard periodogram for certain observing cadences and selection functions. \citet{Zechmeister09} detail the required modifications to the harmonic formalism of \eq{LombScargle} to allow the mean to float in the model. In the matrix formalism, the modification is much more straightforward: all that is required is to add a column of ones to the $X_\omega$ matrix before computing the power via \eq{LombScargle2}. This column of ones corresponds to a third entry in the parameter vector $\theta$, and acts as a uniform constant offset for all data points.

For well-sampled data, there is usually very little difference between a standard periodogram and a floating-mean periodogram. Where this becomes important is if selection effects or observing cadences cause there to be preferentially more observations at certain phases of the light curve: a toy example demonstrating this situation is shown in \fig{floating_mean}. The data are drawn from a sinusoid with Gaussian errors, and data with a magnitude fainter than 16 are removed to simulate an observational bias (left panel). Because of this observational bias, the mean of the observed data are a poor predictor of the true mean, causing the standard fixed-mean method to poorly fit the data and miss the input period (upper-right panel). The floating-mean approach is able to automatically adjust for this bias, resulting in a periodogram which readily detects the input period of 0.622 days (lower-right panel).

\subsection{Truncated Fourier Models}
\sectlabel{multiterm}

\begin{figure}
  \centering
  \includegraphics[width=\textwidth]{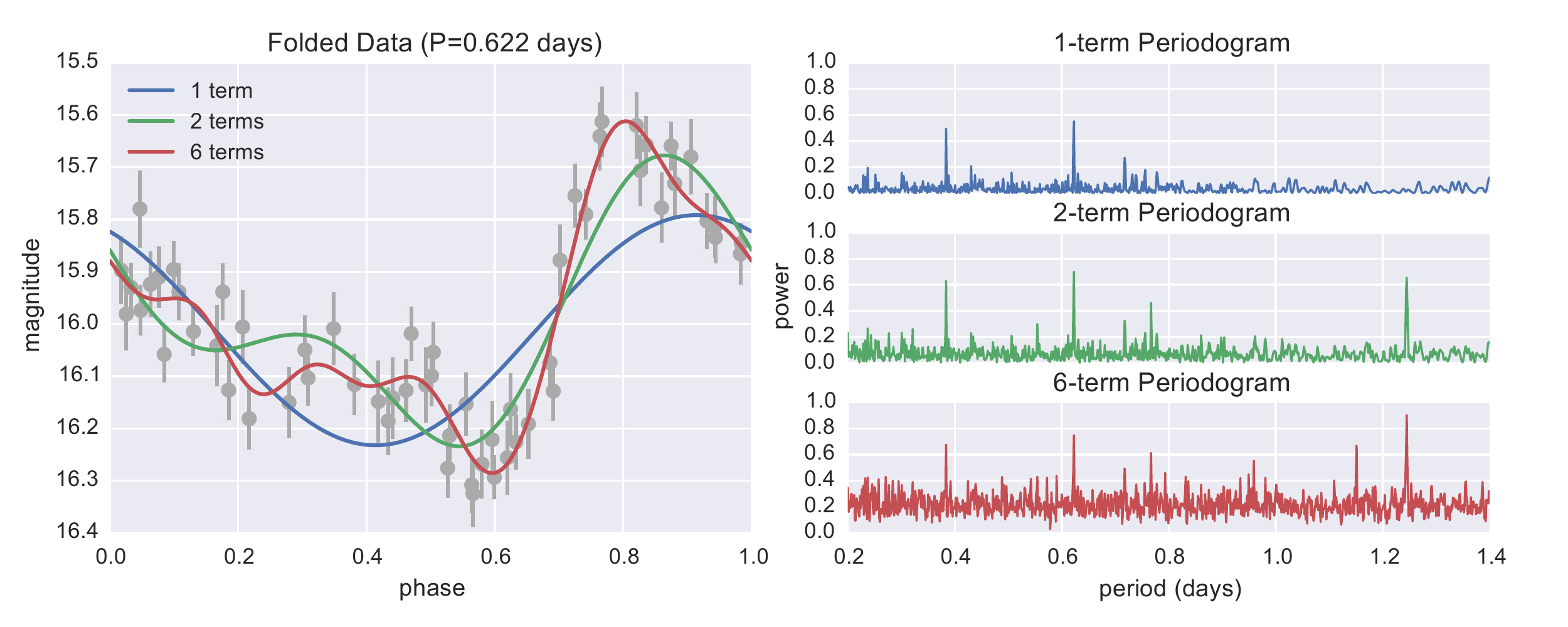}
  \caption{
    The model fits and periodograms for several truncated Fourier models.
    The data are the same as those in \fig{basic_example}. Note that the
    higher-order models will generally show periodogram peaks at multiples
    of the true fundamental frequency $P_0$: this is because for integer $n$
    less than the number of Fourier terms in the model, $P_0$ is a higher
    harmonic of the model at $P=nP_0$. Additionally, the increased degrees of
    freedom in the higher-order models let them fit better at any frequency,
    which drives up the ``background'' level in the periodogram.
  }
  \figlabel{multiterm_example}
\end{figure}

As mentioned above, the standard periodogram is equivalent to fitting a single-term stationary sinusoidal model to the data. A natural extension is to instead use a multiple-term sinusoidal model, with frequencies at integer multiples of the fundamental frequency. With $N$ Fourier terms, there are $2N + 1$ free parameters, and the model is given by
\begin{equation}
  y(t|\omega,\theta) = \theta_0 + \sum_{n=1}^N \left[\theta_{2n - 1}\sin(n\omega t) + \theta_{2n}\cos(n\omega t)\right].
\end{equation}
Because this model remains linear in the parameters $\theta$, it can be easily accommodated into the matrix formalism above. For example, an $N = 2$-term floating-mean model can be constructed by building a design matrix $X_\omega$ with $2N + 1 = 5$ columns:
\begin{equation}
X_\omega^{(2)} = \left[
\begin{array}{ccccc}
1 & \sin(\omega t_1) & \cos(\omega t_1) & \sin(2\omega t_1) & \cos(2\omega t_1)\\
1 & \sin(\omega t_2) & \cos(\omega t_2) & \sin(2\omega t_2) & \cos(2\omega t_2)\\
1 & \sin(\omega t_3) & \cos(\omega t_3) & \sin(2\omega t_3) & \cos(2\omega t_3)\\
\vdots & \vdots & \vdots & \vdots & \vdots \\
1 & \sin(\omega t_N) & \cos(\omega t_N) & \sin(2\omega t_N) & \cos(2\omega t_N)\\
\end{array}
\right]
\end{equation}
Computing the power via \eq{LombScargle2} using $X_\omega^{(2)}$ will give the two-term periodogram. For larger $N$, more columns are added, but the periodogram can be computed in the same manner. \fig{multiterm_example} shows a few examples of this multiterm Fourier approach as applied to the simulated RR Lyrae light curve from \fig{basic_example}, and illustrates several important insights into the subtleties of this type of multiterm fit.

First, we see in the right panel that all three models show a clear signal at the true period of $P_0 = 0.622$ days. The higher-order models, however, also show a a spike in power at $P_1 = 2 P_0$: the reason for this is that for and $N>1$-term model, the period $P_0$ is the first harmonic of a model with fundamental frequency $2P_0$, and the higher-order models contain the single-period result.

Second, notice that as the number of terms is increased, the general ``background'' level of the periodogram increases. This is due to the fact that the periodogram power is inversely related to the $\chi^2$ of the fit at each frequency. A more flexible higher-order model can better fit the data at all periods, not just the true period. Thus in general the observed power of a higher-order Fourier model will be everywhere higher than the power of a lower-order Fourier model.

\subsection{Regularized Models}
\sectlabel{regularization}

\begin{figure}
  \centering
  \includegraphics[width=\textwidth]{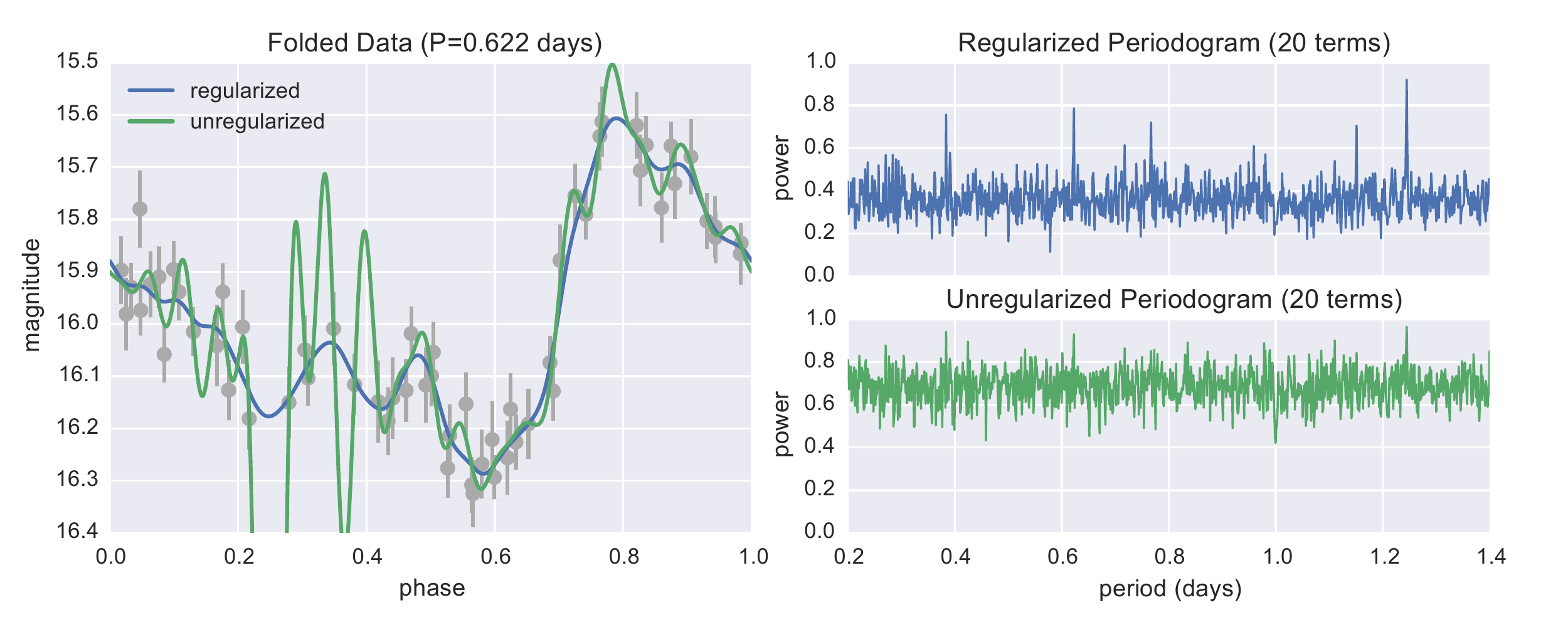}
  \caption{
    The effect of regularization on a high-order model. The data is the same as
    those in \fig{basic_example}. We fit a 20-term truncated Fourier model to
    the data, with and without a regularization term. Without regularization,
    the model oscillates widely to fit the noise in the data. The
    regularization term effectively damps the higher-order Fourier modes and
    removes this oscillating behavior, leading to a more robust model with
    stronger periodogram peaks.
  }
  \figlabel{regularized_example}
\end{figure}

The previous sections raise the question: how complicated a model should we use? We have seen that as we add more terms to the fit, the model will come closer and closer to the observed data.  In the extreme case, when the number of model parameters equals the number of data points, the model can fit the data {\it exactly} regardless of frequency and the periodogram will be everywhere unity (though in most cases, numerical inaccuracies prevent a truly perfect fit). For very high-order models, the fit becomes very sensitive to the noise in the data, and we say we have {\it over-fit} the data. This can be addressed by explicitly truncating the series, but we can also use a {\it regularization} term to mathematically enforce a less complicated model.

A regularization term is an explicit penalty on the magnitude of the model parameters $\theta$, and can take a number of forms. For computational simplicity here we'll use an {\it L2 regularization} -- also known as Tikhonov Regularization \citep{Tikhonov1963} or Ridge Regression \citep{Hoerl1970} -- which is a quadratic penalty term in the model parameters added to the $\chi^2$. Mathematically, this is equivalent in the Bayesian framework to using a zero-mean Gaussian prior on the model parameters.

We encode our regularization in the matrix $\Lambda = {\rm diag}([\lambda_1, \lambda_2 \cdots \lambda_M])$ for a model with $M$ parameters, and construct a ``regularized'' $\chi^2$:
\begin{equation}
  \eqlabel{chi2reg}
  \chi_\Lambda^2(\omega) = (y - X_\omega\theta)^T\Sigma^{-1}(y - X_\omega\theta) + \theta^T\Lambda\theta
\end{equation}
Minimizing this regularized $\chi^2$, solving for $\theta$, and plugging into the expression for $P_N$ gives us the regularized counterpart of \eq{LombScargle2}:
\begin{equation}
  \eqlabel{LombScargleReg}
  P_{N,\Lambda}(\omega) = \frac{y^T\Sigma^{-1}X_\omega~[X_\omega^T\Sigma^{-1}X_\omega + \Lambda]^{-1}~X_\omega^T\Sigma^{-1}y}{y^T\Sigma^{-1}y}.
\end{equation}
Notice that the effect of this regularization term is to add a diagonal penalty to the normal matrix $X_\omega^T\Sigma^{-1}X_\omega$, which has the additional feature that it can correct ill-posed models where the normal matrix is non-invertible. This feature of the regularization will become important for the multiband models discussed below.

In \fig{regularized_example}, we compare a regularized and unregularized 20-term truncated Fourier model on our simulated RR Lyrae light curve. We use $\lambda = 0$ on the offset term, and make the penalty $\lambda_j$ progressively larger for each harmonic component. The result of the regularized model is less over-fitting to the input data, and stronger periodogram peaks.

\section{A Multiple-Band Model}
\sectlabel{multiband}
In this section we will apply what was learned in the previous sections to construct the {\it multiband periodogram} which can flexibly account for heterogeneous sources of data for a single object.
To compute such a periodogram, we will take advantage of the easy extensibility of the matrix formalism which led to our generalizations above.
The multiband model contains the following features:
\begin{enumerate}
  \item An $N_{base}$-term truncated Fourier fit which models a latent parameter, which we'll call the ``overall variability''.
  \item A set of $N_{band}$-term truncated Fourier fits, each of which models the residual of a single band from this overall variability.
\end{enumerate}
The total number of parameters for $K$ filters is then $M_K = (2N_{base} + 1) + K(2N_{band} + 1)$. As a result, for each band $k$ we have the following model of the observed magnitudes:
\begin{eqnarray}
  \eqlabel{multiband_model}
  y_k(t|\omega,\theta) = &\theta_0 + \sum_{n=1}^{M_{base}} \left[\theta_{2n - 1}\sin(n\omega t) + \theta_{2n}\cos(n\omega t)\right]& +\nonumber\\ 
  &\theta^{(k)}_0 + \sum_{n=1}^{M_{band}} \left[\theta^{(k)}_{2n - 1}\sin(n\omega t) + \theta^{(k)}_{2n}\cos(n\omega t)\right].&
\end{eqnarray}
The important feature of this model is that {\it all bands} share the same base parameters $\theta$, while their offsets $\theta^{(k)}$ are determined individually.

We can construct the normalized periodogram for this model by building a sparse design matrix with $M_K$ columns. Each row corresponds to a single observation through a single band. Columns corresponding to the base model and the matching observation band will have nonzero entries; all other columns will be filled with zeros. For example, the $N_{base}=1$ and $N_{band}=0$ model corresponds to one with a simple single-term periodic base frequency, and an independent constant offset term in each band. The associated design matrix depends on the particular data, but will look similar to this:
\begin{equation}
X_\omega^{(1,0)} = \left[
\begin{array}{cccccccc}
1 & \sin(\omega t_1) & \cos(\omega t_1) & 1 & 0 & 0 & 0 & 0\\
1 & \sin(\omega t_2) & \cos(\omega t_2) & 0 & 0 & 0 & 0 & 1\\
1 & \sin(\omega t_3) & \cos(\omega t_3) & 0 & 0 & 0 & 1 & 0\\
\vdots & \vdots & \vdots & & & \vdots & &\\
1 & \sin(\omega t_N) & \cos(\omega t_N) & 0 & 0 & 1 & 0 & 0\\
\end{array}
\right]
\end{equation}
Here the nonzero entries of the final five columns are binary flags indicating the $(u, g, r, i, z)$-band of the given observation: for this example, the first row is a $u$-band measurement, the second is a $z$-band, the third is a $i$-band, etc., as indicated by the position of the nonzero matrix element within the row.

On examination of the above matrix, it's clear that the columns are not linearly independent (i.e. $X_\omega$ is low-rank), and thus the parameters of the best-fit model will be degenerate. Intuitively, this is due to the fact that if we add an overall offset to the base model, this can be perfectly accounted for by subtracting that same offset from each residual. Mathematically, the result of this is that the normal matrix $X_\omega^T\Sigma^{-1}X_\omega$ will be non-invertible, and thus the periodogram is ill-defined. In order to proceed, then, we'll either have to use a different model, or use a cleverly-constructed regularization term on one of the offending parameters.

We'll choose the latter here, and regularize all the band columns while leaving the base columns un-regularized: for the above $X_\omega$ matrix, this regularization will look like
\begin{equation}
  \Lambda^{(1,0)} = {\rm diag}([0, 0, 0, \epsilon, \epsilon, \epsilon, \epsilon, \epsilon])
\end{equation}
where $\epsilon$ is some small fraction of the trace of the normal matrix $[X_\omega^T\Sigma^{-1}X_\omega]$. The logic of this choice of regularization is that any component of the model which is common to all bands will be preferentially reflected in the base terms, with independent behavior reflected in the individual band terms. Setting $\epsilon$ to some small fraction of the trace ensures that the regularization effect on the remaining model will be small. With this regularization in place, the model is well-posed and \eq{LombScargleReg} can be used to straightforwardly compute the power. The effective number of free parameters for such a regularized $(N_{base}, N_{band})$ model with $K$ filters is
$M_K^{eff} = 2N_{base}^{eff} + K(2N_{band} + 1)$ where $N_{base}^{eff} = \max(0, N_{base} - N_{band})$ is the effective number of base terms.

The final remaining piece to mention is our assumption in \eq{ycentered} that the data are centered. This is required so that the simple form of the reference $\chi^2_0$ remains valid. For the multiband model, this assumption requires that the data satisfy \eq{ycentered} {\it within each band}: equivalently, we could lift this assumption and compute the reference $\chi^2_0$ of the multiband model with an independent floating mean within each band; the results will be identical.

This multiband approach, then, actually comprises a set of models indexed by their value of $N_{base}$ and $N_{band}$. The most fundamental models have $(N_{base}, N_{band}) = (1,0)$ and $(0,1)$, which we'll call the {\it shared-phase} and {\it multi-phase} models respectively. In the shared-phase model, all variability is assumed to be shared between the bands, with only the fixed offset between them allowed to float. In the multi-phase model, each band has independent variability around a shared fixed offset.

\subsection{Relationship of Multiband and Single-band approaches}
\sectlabel{relationship}
The {\it multi-phase} $(N_{base}=0, N_{band}=1)$ model turns out to be a particularly special case.
Here the base model is a simple global offset which is degenerate with the offsets in each band, so that the design matrix $X_\omega$ can be straightforwardly rearranged as block-diagonal.
A block-diagonal design matrix in a linear model indicates that components of the model are being solved independently: here these independent components amount to the single-band floating-mean model from \sect{floating_mean}, fit independently for each of the $K$ bands.
This particular multiband model can give us insight into the relationship between single-band and multiband approaches.

For band $k$, we'll denote the single-band floating-mean periodogram as
\begin{equation}
  P_N^{(k)}(\omega) = 1 - \frac{\chi^2_{min, k}(\omega)}{\chi^2_{0,k}}
\end{equation}
The full multiband periodogram is given by
\begin{equation}
  P_N^{(0,1)}(\omega) = 1 - \frac{\sum_{k=1}^K\chi^2_{min, k}(\omega)}{\sum_{k=1}^K\chi^2_{0,k}}
\end{equation}
and it can easily be shown that the $P_N$ can be constructed as a weighted sum of $P_N^{(k)}$:
\begin{equation}
  P_N^{(0,1)}(\omega) = \frac{\sum_{k=1}^K\chi^2_{0,k}P_N^{(k)}}{\sum_{k=1}^K\chi^2_{0,k}}.
\end{equation}
We see that this particular multiband periodogram is identical to a weighted sum of standard periodograms in each band, where the weights $\chi^2_{0,k}$ are a reflection of both the number of measurements in each band, and how much those measurements deviate from a simple constant reference model.

\subsection{Multiband Periodogram for Simulated Data}
\sectlabel{Simulated}

\begin{figure}
  \centering
  \includegraphics[width=\textwidth]{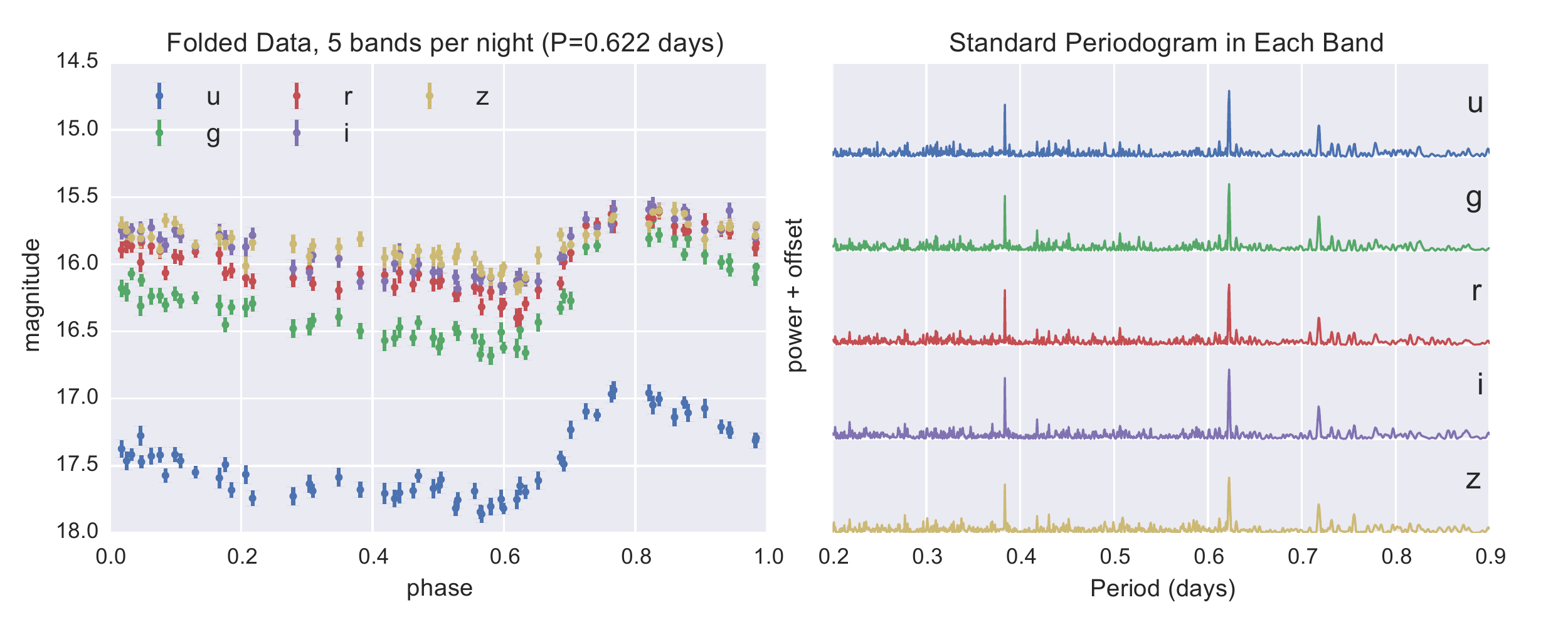}
  \includegraphics[width=\textwidth]{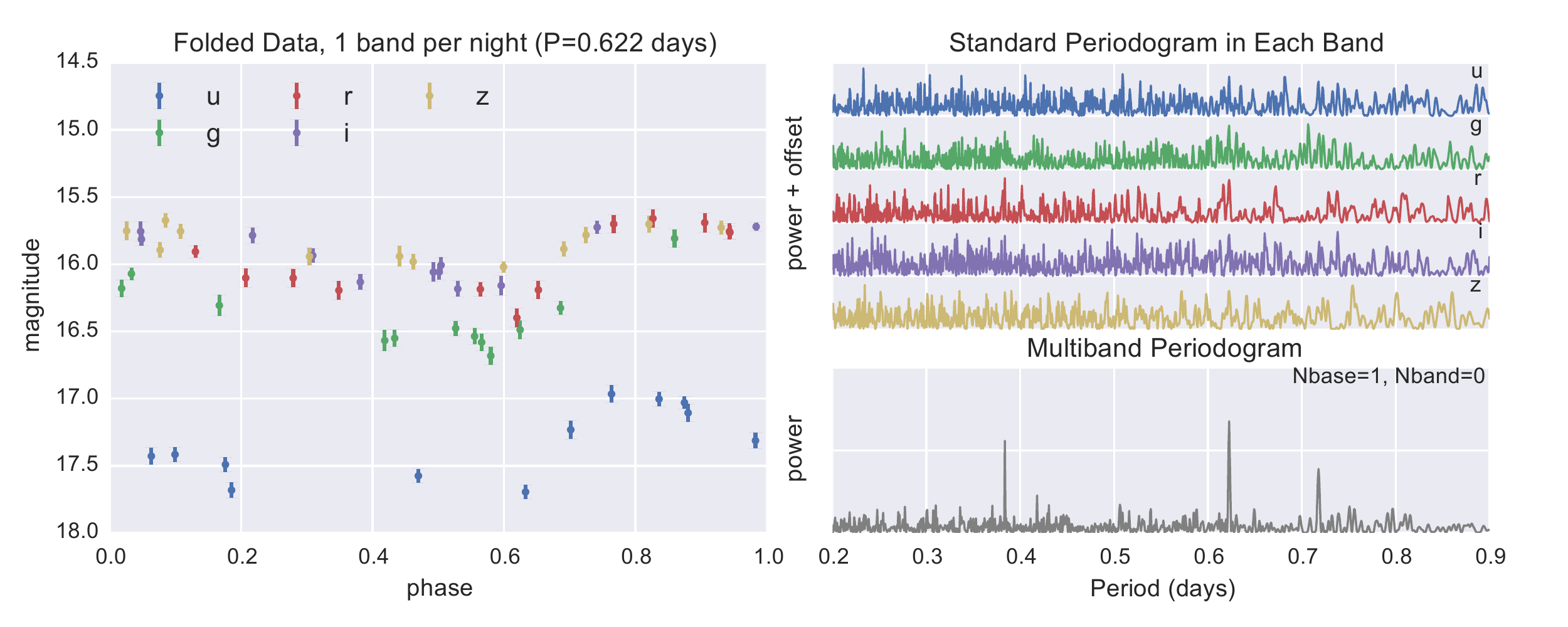}
  \caption{
    An illustration of the performance of the multiband periodogram. The
    upper panels show simulated $ugriz$ observations of an RR Lyrae light
    curve in which all 5 bands are observed each night. With 60 observations
    in each band, a periodogram computed from any single band is sufficient to
    determine the true period of 0.622 days. The lower panels show the same
    data, except with only a single $ugriz$ band observed each night (i.e.
    12 observations per band). In this case, no single band has enough
    information to detect the period. The shared-phase multiband approach
    of \sect{multiband} (lower-right panel) combines the information from
    all five bands, and results in a significant detection of the true period.
  }
  \figlabel{multiband_sim}
\end{figure}

\begin{figure}
  \centering
  \includegraphics[width=0.8\textwidth]{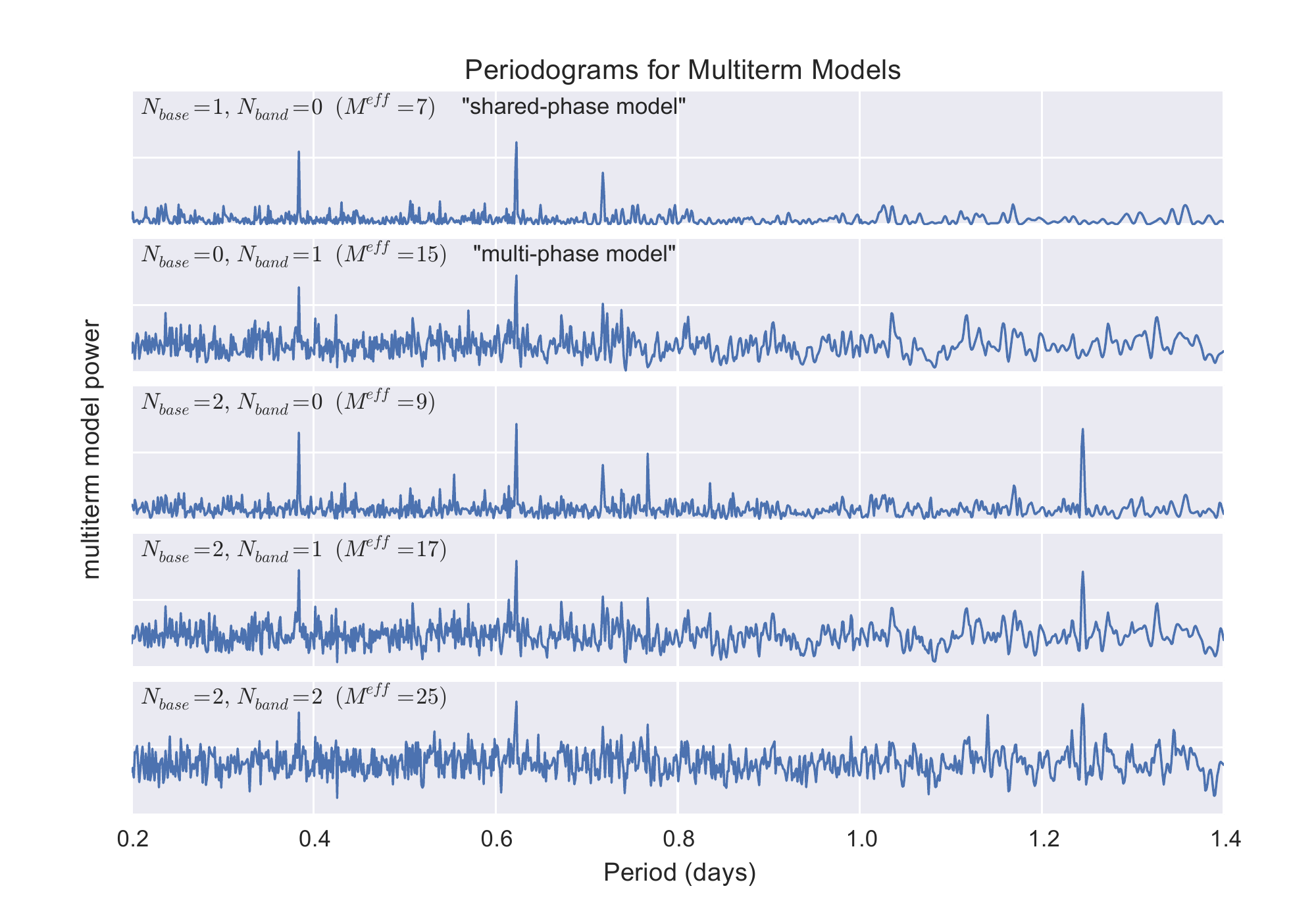}
  \caption{
    Comparison of the periodograms produced by various multiband models.
    The data is the same as that used in \fig{multiband_sim}. $N_{base}$ gives
    the number of Fourier terms in the base model, and $N_{band}$ gives the
    number of Fourier terms used to fit the residuals around this model within
    each band. The characteristics discussed with previous figures are also
    seen here: in particular, the level of ``background noise'' in the
    periodogram grows with the model complexity $M$,
  } 
  \figlabel{multiband_models}
\end{figure}

Before applying the multiband method to real data, we will here explore its effectiveness on a simulated RR Lyrae lightcurve.
The upper panels of \fig{multiband_sim} show a multiband version of the simulated RR Lyrae light curve from \fig{basic_example}.
The upper-left panel shows 60 nights of observations spread over a 6-month period, and for each night all five bands ({\it u,g,r,i,z}) are recorded.
Using the typical approach from the literature, we individually compute the standard normalized periodogram within each band: the results are shown in the upper-right panel.
The data are well-enough sampled that a distinct period of 0.622 days can be recognized within each individual band, up to the aliasing effect discussed in \sect{simple_period}.
Previous studies have made use of the information in multiple bands to choose between aliases and estimate uncertainties in determined periods \citep[e.g.][]{Oluseyi12,Sesar2010}.
While this approach is sufficient for well-sampled data, it becomes problematic when the multiband data are sparsely sampled.

The lower panels of \fig{multiband_sim} show the same 60 nights of data, except with only a {\it single} band observation recorded each night.
The lower-left panel shows the observations as a function of phase, and the lower-right panels show the periodograms derived from the data.
With only 12 observations for each individual band, it is clear that there is not enough data to accurately determine the period within each single band. The shared-phase multiband approach (i.e. $N_{base}=1,N_{band}=0$), shown in the lower-right panel, fits a single model to the full data and clearly recovers the true frequency of 0.622 days.

This shared-phase $(1,0)$ model is only one of the possible multiband options, however: \fig{multiband_models} shows multiband fits to this data for models with various choices of $(N_{base},N_{band})$.
We see here many of the characteristics noted above for single-band models: as discussed in \sect{multiterm}, increasing the number of Fourier terms leads to power at multiples of the fundamental period, and increased model complexity (roughly indexed by the effective number of free parameters $M^{eff}$) tends to increase the background level of the periodogram, obscuring significant peaks.
For this reason, models with $N_{base} > N_{band}$ are the most promising: they allow a flexible fit with minimal model complexity. Motivated by this, in the next section we'll apply the simplest of this class of models, the $(1, 0)$ shared-phase model, to data from the Stripe 82 of the Sloan Digital Sky Survey.

\section{Application to Stripe 82 RR Lyrae}
\sectlabel{stripe82}
Stripe 82 is a three hundred square degree equatorial region of the sky which was repeatedly imaged through multiple band-passes during phase II of the Sloan Digital Sky Survey \citep[SDSS II, see][]{Sesar2007}.
Here we consider the SDSS II observations of 483 RR Lyrae stars compiled and studied by S10, in which periods for these stars were determined based on empirically-derived light curve templates.
Because the template-fitting method is extremely computationally intensive, S10 first determined candidate periods by taking the top 5 results of the Supersmoother \citep{Reimann94} algorithm applied to the $g$-band; template fits were then performed at each candidate period and the period with the best template fit was reported as the true period. In this section, we make use of this dataset to quantitatively evaluate the effectiveness of the multiband periodogram approach.

\subsection{Densely-sampled Multiband Data}

\begin{figure}
  \centering
  \includegraphics[width=\textwidth]{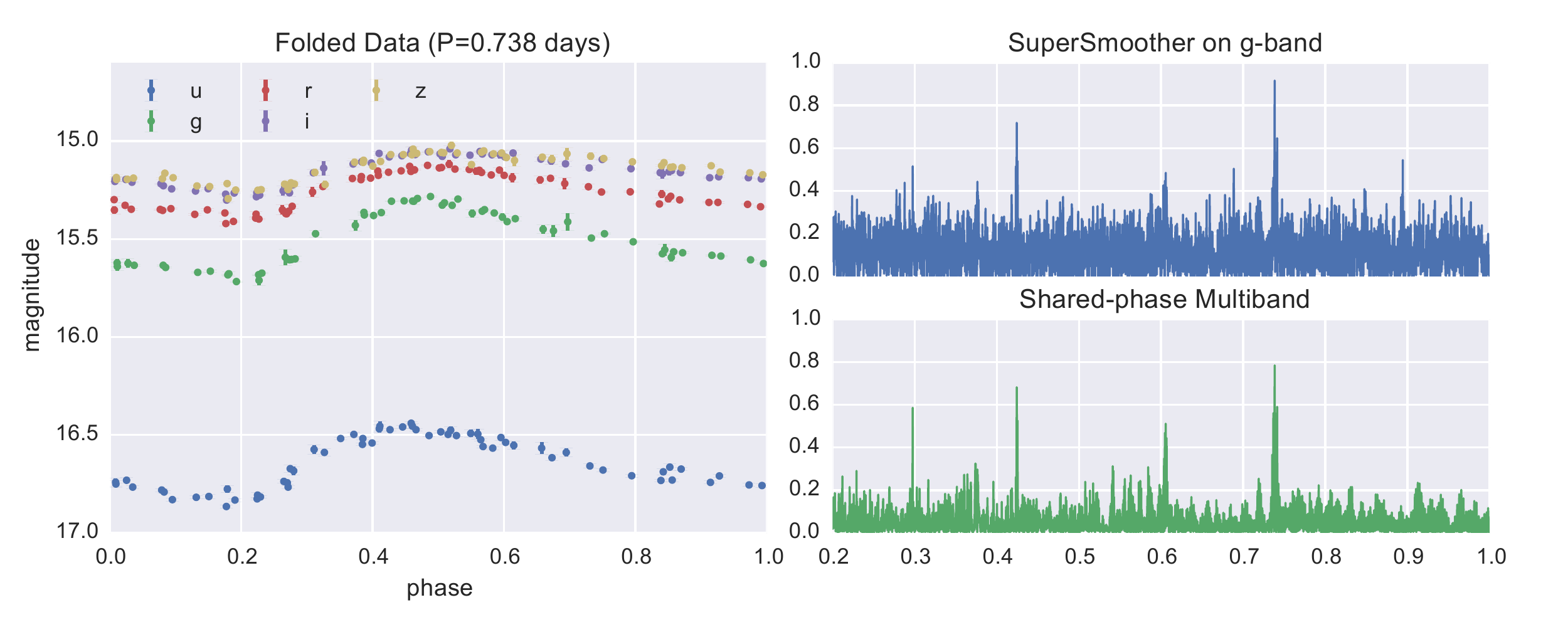}
  \includegraphics[width=\textwidth]{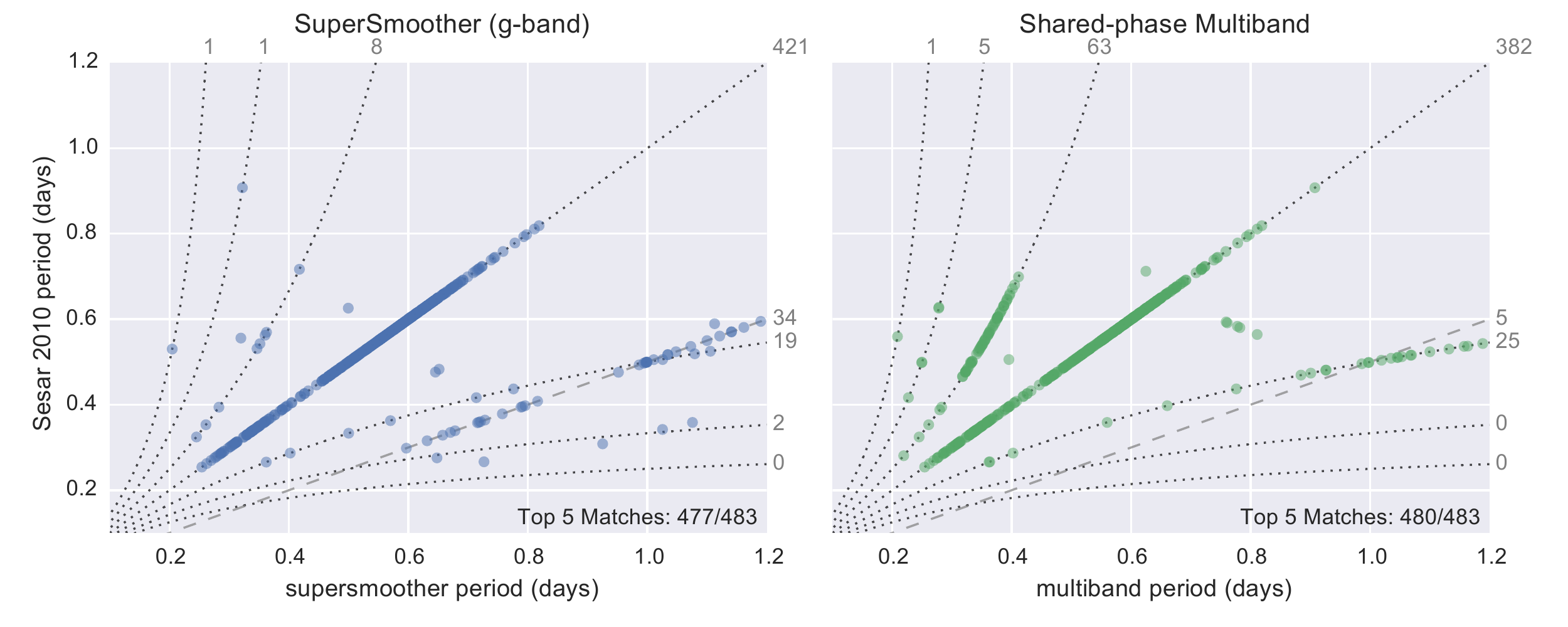}
  \caption{
    Comparison of the Multiband algorithm and single-band supersmoother algorithm on 483 well-sampled RR Lyrae light curves from Stripe 82.
    The upper panels show a representative lightcurve and periodogram fits, while the bottom panels compare the derived periods to the template-based periods reported in S10.
    Shown for reference are the beat aliases (dotted lines) and the multiplicative alias (dashed lines): numbers along the top and right edges of the panels indicate the number of points aligned with each trend.
    The single-band supersmoother model tends to err toward multiplicative aliases, while the multiband model tends to err toward beat frequency aliases.
    Both methods find the correct period among the top 5 significant peaks around 99\% of the time.
  } 
  \figlabel{compare_periods}
\end{figure}

The full S10 RR Lyrae dataset consists of 483 objects with an average of 55 observations in each of the five SDSS $ugriz$ bands spread over just under ten years.  In the upper panels of \fig{compare_periods} we show the observed data for one of these objects, along with the periodogram derived with the single-band supersmoother model and the shared-phase $(0, 1)$-multiband model\footnote{The supersmoother ``periodogram'' $P_{SS}$ is constructed from the minimum sum of weighted model residuals $\bar{r}_{min}$ in analogy with \eq{chi2PN}: $P_{SS}(\omega) = 1 - \bar{r}_{min}(\omega) / \bar{r}_0$, where $\bar{r}_0$ is the mean absolute residual around a constant model.}. Here we have a case which is analogous to that shown for simulated data in the top panels of \fig{multiband_sim}: each band has enough data to easily locate candidate peaks, the best of which is selected via the S10 template-fitting procedure.

The lower panels of \fig{compare_periods} compare the S10 period with the best periods obtained from the 1-band supersmoother (lower-left) and from the shared-phase multiband model (lower-right). To guide the eye, the figure includes indicators of the locations of beat aliases (dotted lines) and multiplicative aliases (dashed lines) of the S10 period.

The best-fit supersmoother period matches the S10 period in 89\% of cases (421/483), while the best-fit multiband period matches the S10 period in 79\% of cases (382/483). The modes of failure are instructive: when the supersmoother model misses the S10 period, it tends to land on a multiplicative alias (i.e. the dashed line). This is due to the flexibility of supersmoother: a doubled period spreads the points out, leading to fewer constraints in each neighborhood and thus a smaller average residual around model. In other words, the SuperSmoother tends to over-fit data which is sparsely-sampled. On the other hand, when the multiband model misses the S10 period, it tends to land on a beat alias between the S10 period and the 1-day observing cadence (i.e. the dotted lines). This is due to the fact that the single-frequency periodic model is biased, and significantly under-fits the data: it cannot distinguish residuals due to underfitting from residuals due to window function effects.

In both models, the S10 period appears among the top 5 periods 99\% of the time: $477/483$ for supersmoother, and $480/483$ for multiband.\footnote{We might expect this correspondence to be 100\% in the case of the $g$-band supersmoother, which was the model used in the first pass of the S10 computation. This discrepancy here is likely due to the slightly different supersmoother implementations used in S10 and in this work. Objects showing this discrepancy are those with very low signal-to-noise.} This suggests that had S10 used the multiband Lomb-Scargle rather than the supersmoother in the first pass for that study, the final results presented there would be for the most part unchanged.

The results of this subsection show that the shared-phase multiband approach is comparable to the supersmoother approach for densely-sampled multiband data, although it has a tendency to get fooled by structure in the survey window. Correction for this based on the estimated window power may alleviate this (see \citet{Roberts87} for an example of such an approach) though in practice selecting from among the top 5 peaks appears to be sufficient.

\subsection{Sparsely-sampled Multiband Data}

\begin{figure}
  \centering
  \includegraphics[width=\textwidth]{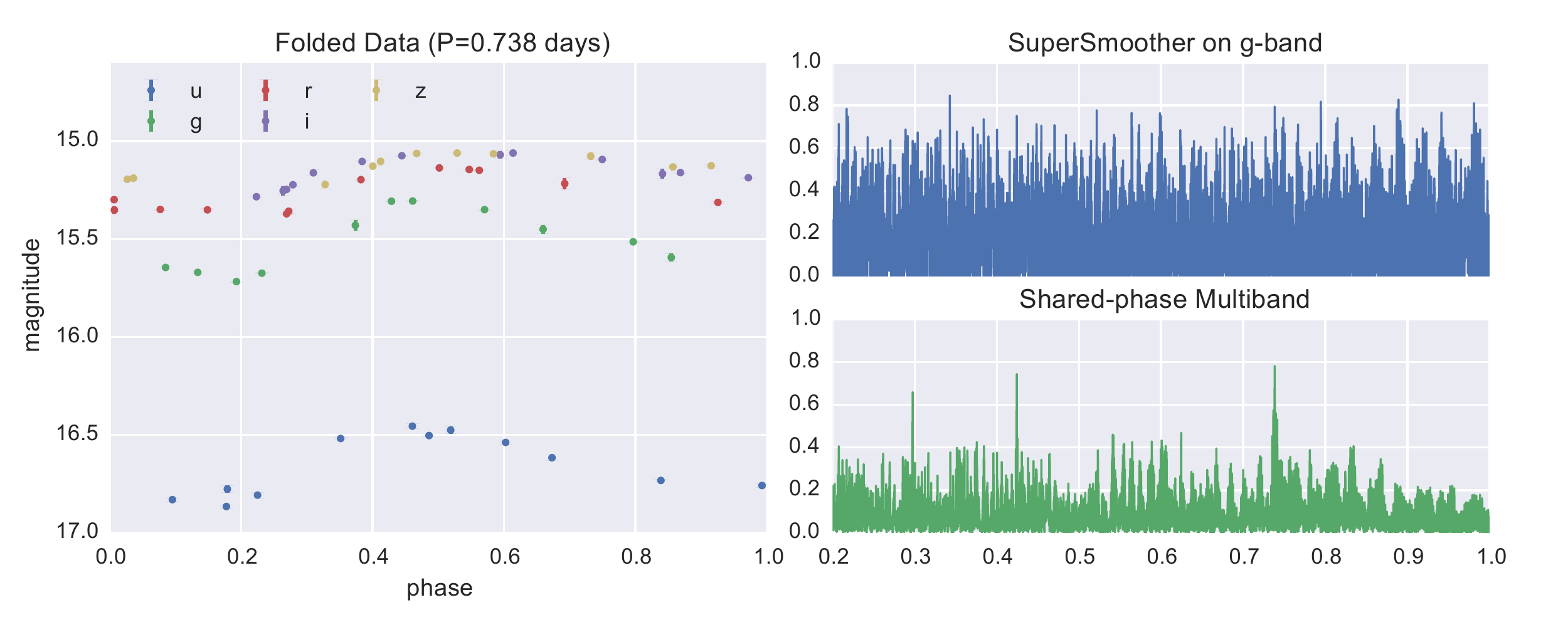}
  \includegraphics[width=\textwidth]{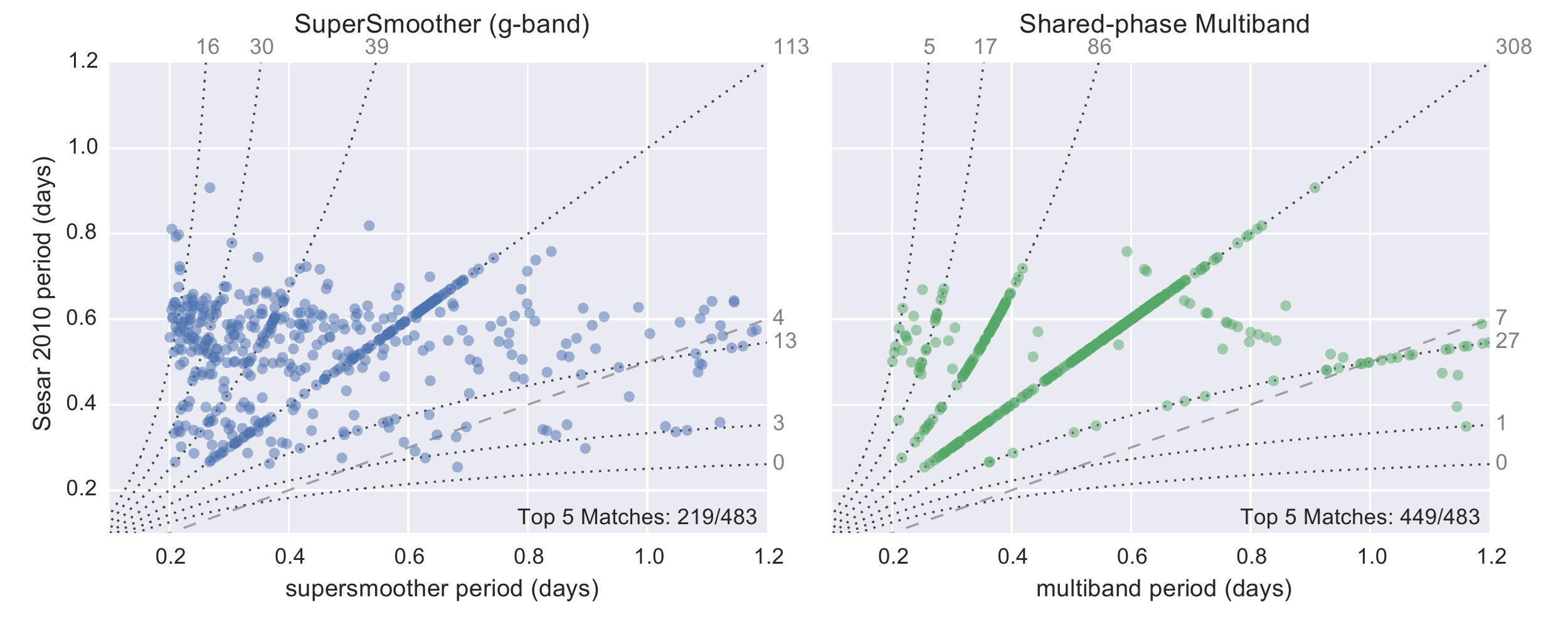}
  \caption{
    This figure repeats the experiment shown in \fig{compare_periods} (see caption there for description), but the data is artificially reduced to only a single-band observation on each evening, a situation reflective of the observing cadence of future large-scale surveys.
    In this case, the single-band SuperSmoother strategy used as a first pass in S10 fails: there is simply not enough data in each band to recover an accurate period estimate. The correct period is among the top 5 candidates in fewer than 50\% of cases.
    The shared-phase multiband approach utilizes information from all five bands, and returns much more robust results: even with the greatly-reduced data, the true period is among the top 5 candidates in 93\% of cases.
  } 
  \figlabel{compare_periods_reduced}
\end{figure}

Above we saw that the multiband model is comparable to methods from the literature for densely-sampled data. Where we expect the multiband approach to gain an advantage is when the data are sparsely sampled, with data through only a single band at each observation time. To simulate this, we reduce the size of the Stripe 82 RR Lyrae dataset by a factor of 5, keeping only a single band of imaging each night: an average of 11 observations of each object per band. This is much closer to the type of data which will be available in future multiband time-domain surveys.

The upper panels of \fig{compare_periods_reduced} show an example light curve from this reduced dataset, along with the supersmoother and multiband periodograms derived from this data. Analogously to the lower panels of \fig{multiband_sim}, the single-band supersmoother model loses the true period within the noise, while the shared-phase multiband model still shows prominent signal near the S10 period.

The lower panels of \fig{compare_periods_reduced} show the relationship between the S10 periods (based on the full dataset) and the periods derived with each model from this reduced dataset. It is clear that the supersmoother model is simply over-fitting noise with this few data points: the top period matches S10 in only 32\% of cases (compared to 87\% with the full dataset), and the top 5 periods contain the S10 period only 45\% of the time. The failure mode is much less predictable as well: rather than being clustered near aliases, most of the period determinations are scattered seemingly randomly around the parameter space.

The multiband method does much better on the reduced dataset. Even with an 80\% reduction in the number of observations, the multiband method matches the S10 period 64\% of the time (compared to 79\% with the full dataset), and the top 5 peaks contain the S10 period 94\% of the time (compared to 99\% with the full dataset). This performance is due to the fact that the multiband algorithm has relatively few parameters, but is yet able to flexibly accommodate data from multiple observing bands. In particular, this suggests that with the multiterm periodogram, the S10 analysis could have been done effectively with only a small fraction of the available data. This bodes well for future surveys, where data on variable stars will be much more sparsely sampled.

\subsection{Potential Improvements to the Multiband Method}
The fact that the multiband model does not select the top frequency each time points to its weaknesses: first, as a Fourier-like model, it tends to respond to frequency structure in the window function as well as frequency structure in the data.
This is a result of the very model simplicity which causes its success in the case of sparse multiband data: it cannot disentangle bias in the model from bias due to the survey window.
This could potentially be accounted for by correcting for the effect of the estimated window function; one potential method for this involves estimating the deconvolution of the window power and the observed power \citep{Roberts87}.
It may also be possible to propose a multiband extension of CARMA \citep{Kelly14}, or another forward-modeling approach to detecting periodicity.

One potentially fruitful avenue of research which we do not study here is the application of other types of regularization to the higher-order periodogram.
In particular, L1 regularization (also known as Lasso regression) could lead to interesting results: L1 regularization is similar in spirit to the Tikhonov regularization discussed in \sect{regularization}, but tends toward sparsity in the model parameters \citep[see, e.g.][for a discussion]{ICVG2014}.
Such an approach could provide a useful tradeoff between model complexity and bias in the case of higher-order truncated Fourier models.

Another potentially interesting extension of the multiband case would be to define and make use of physically-motivated priors in the light-curve shape.
This approach could allow the model bias to be decreased without a commensurate increase in model complexity, which is what causes poor performance in the case of sparsely-sampled noisy data.
As an example of such a physically-motivated prior, consider that the paths of RR Lyrae stars through color-color and color-magnitude space are constrained by known astrophysical processes in the structure of the stars \citep[e.g., see Fig. 5 in][]{Szabo2014}. Making use of this information could help break degeneracies in period determination with higher-order models.

\section{Prospects for Multiband Periodograms with LSST}
\sectlabel{LSST}
Previously, \citet{Oluseyi12} evaluated the prospects of period finding in early LSST data, and found results which were not encouraging.
Using the conservative criterion of a 2/3 majority among the top single-band supersmoother periods in the $g$, $r$, and $i$ bands, they showed that, depending on spectral type, finding reliable periods for the brightest ($g \sim 20$) RR Lyrae stars will require several years of LSST data, while periods for some of the faintest ($g \sim 25$) stars will not be reliable with even ten years of data!

One potential remedy is to move away from general models like supersmoother and lomb-scargle to specific template-fitting methods such as those used in S10.
Indeed, such methods perform well even for sparsely-sampled multiband data such as those from the PanSTARRS survey; the primary drawback is that such blind template fits are computationally extremely expensive: they involve nonlinear optimizations over each of several hundred candidate templates at each of tens of thousands of candidate frequencies (B. Sesar, private communication).
Thus the template-fitting method, though it produces accurate periods, in practice requires several hours of CPU time for a well-sampled period grid for a single source (compared to several seconds for the multiband periodogram proposed here).
Note that several hours per object is orders-of-magnitude too slow in the case of LSST; to estimate periods for a billion stars on a 1000-core machine in a year requires a compute-time budget of only 30 seconds per light curve.

Because of the computational expense of the pure template-fitting method, when working with SDSS II data S10 performed a first-pass with a single-band supersmoother to establish candidate periods, which were in turn evaluated with template-fitting approach.
Here we show that such a hybrid strategy combining the multiband periodogram and the S10 template fits will be useful for determining periodicity of variables in early LSST data releases, greatly improving on the outlook presented in \citet{Oluseyi12}.

We suggest the following procedure for determining periods in future multiband datasets:
\begin{enumerate}
   \item As a first pass, find a set of candidate frequencies using the multiband periodogram. This is a fast linear optimization that can be trivially parallelized.
   \item Within these candidate frequencies, use the more costly template-fitting procedure to choose the optimal period from among the handful of candidates.
   \item Compute a goodness-of-fit statistic for the best-fit template to determine whether the fit is suitable; if not, then apply the template-fitting procedure across the full period range.
\end{enumerate}
Here we briefly explore simulated LSST observations of RR Lyrae stars in order to gauge the effectiveness of the first step in this strategy; the effectiveness of the template-fitting step will be explored further in future work.
Rather than doing the full analysis including the final template fits, we will focus on the ability of the multiband periodogram to quickly provide suitable candidate periods under the assumption that the S10 template algorithm will then select or reject the optimal period from this set.

\subsection{LSST Simulations}

We use a simulated LSST cadence \citep{opsim1, opsim2, opsim3} in 25 arbitrarily chosen fields
that are representative of the anticipated main survey temporal coverage.
We simulate a set of 50 RR Lyrae observations with the S10 templates, with a range of apparent magnitudes between 
$g=20$ and $g=24.5$, corresponding to bright-to-faint range of LSST main-survey observations, and with expected 
photometric errors computed using eqs.~4--6  from \citet{Ivezic08LSST}. 
Given the capability of template-fitting to choose among candidate periods, we use a more relaxed period-matching criterion than in \citet{Oluseyi12}: when evaluating the single-band supersmoother, we require that the true period is among the five periods determined independently in the $u, g, r, i, z$ bands; in the multiband case we require that the true period is among the top five peaks in the multiband periodogram.

\fig{LSST_sims} shows the fraction of stars where this period matching criterion is met as a function of $g$-band magnitude and subset of LSST data.
The solid lines show the multiband results; the dashed lines show the single-band supersmoother results; and the shading helps guide the eye for the sake of comparison.
Because of our relaxed matching criteria, even the single-band supersmoother results here are much more optimistic than the \citet{Oluseyi12} results (compare to Figure 15 in that work): the supersmoother result here can be considered representative of a best-case scenario for \adhoc{} single-band fits.
Without fail, the multiband result exceeds this best-case single-band result; the improvement is most apparent for faint stars, where the greater model flexibility of the supersmoother causes it to over-fit the noisy data.

The performance of the multiband periodogram points to much more promising prospects for science with variable stars than previously reported.
In particular, even with only six months of LSST data, we can expect to correctly identify the periods for over 60\% of stars brighter than $g=22$; with the first two years of LSST observations, this increases to nearly 100\%; with five years of data, the multiband method identifies the correct period for 100\% of even the faintest stars.
Part of this improvement is due to the performance of the shared-phase multiband model with noisy data, and part of this improvement is due to the relaxed period-matching constraints enabled by the hybrid approach of periodogram-based and template-based period determination.



\begin{figure}
  \centering
  \includegraphics[width=0.7\textwidth]{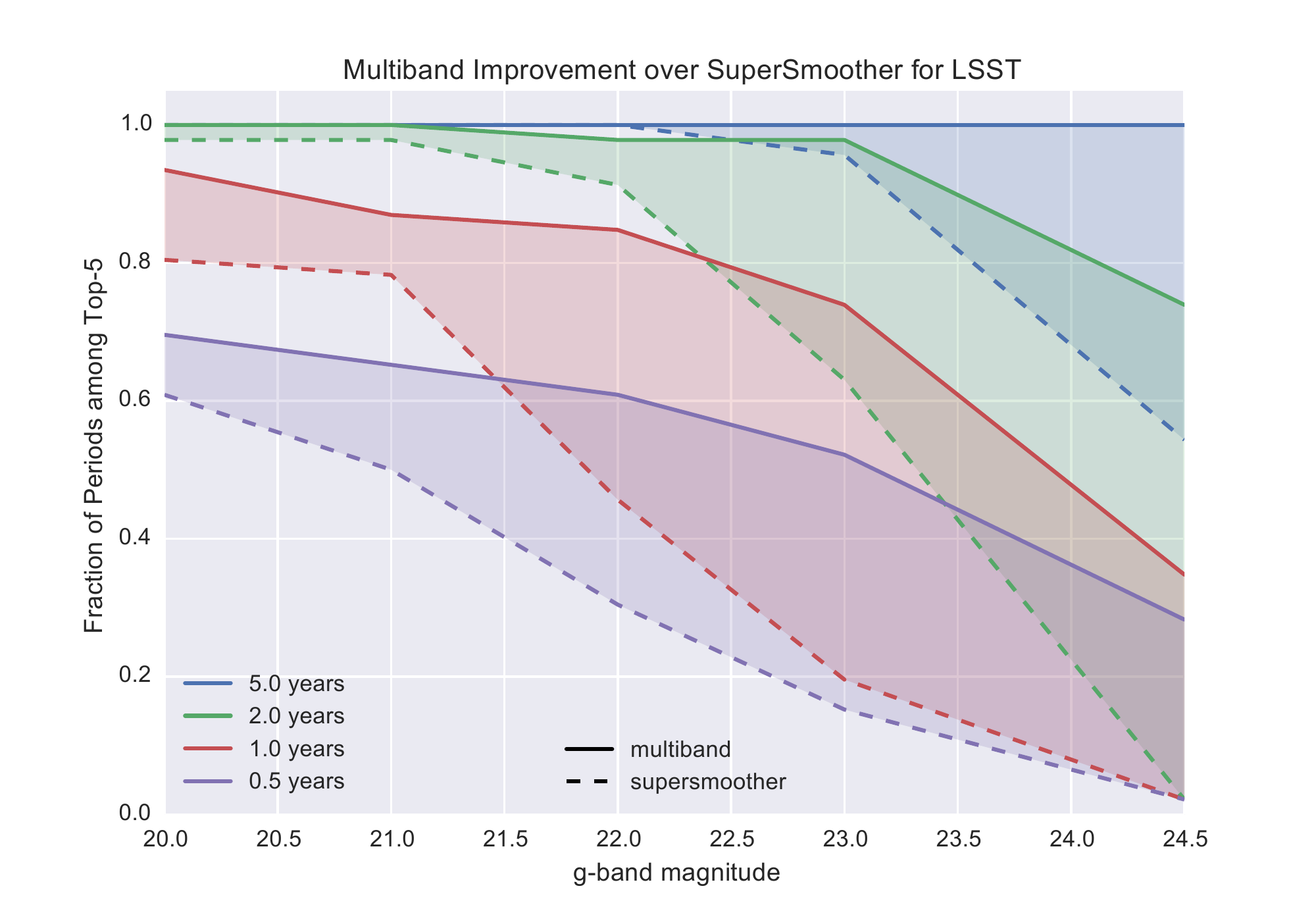}
  \caption{
    Fraction of periods correctly determined for LSST RR Lyrae as a function
    of the length of the observing season and the mean $g$-band magnitude, for the multiband (solid lines)
    and single-band supersmoother (dashed lines) approaches.
    The multiband method is superior to the single-band supersmoother approach in all cases, and especially for the faintest objects.
  } 
  \figlabel{LSST_sims}
\end{figure}

\section{Discussion and Conclusion}
\sectlabel{discussion}

We have motivated and derived a multiband version of the classic Lomb-Scargle method for detecting periodicity in astronomical time-series.
Experiments on several hundred RR Lyrae stars from the SDSS Stripe 82 dataset indicate that this method outperforms methods used previously in the literature, especially for sparsely-sampled light curves with only single bands observed each night.
While there are potential areas of improvement involving corrections to window function artifacts and accounting for physically-motivated priors, the straightforward multiband model outperforms previous \adhoc{} approaches to multiband data.

Looking forward to future variable star catalogs from PanSTARRS, DES, and LSST, there are two important constraints that any analysis method must meet: the methods must be able to cope with heterogeneous and noisy observations through multiple band-passes, and the methods must be fast enough to be computable on millions or even billions of objects.
The multiband method, through its combination of flexibility and model simplicity, meets the first constraint: as shown above, in the case of sparsely-sampled noisy multiband data, it out-performs previous approaches to period determination.
It also meets the second constraint: it requires the solution of a simple linear model at each frequency, compared to a rank-based sliding-window model in the case of supersmoother, a nonlinear optimization in the case of template-fitting, and a Markov Chain Monte Carlo analysis in the case of CARMA models.
In our own benchmarks, we found the multiband method to be several times faster than the single-band supersmoother approach, and several orders of magnitude faster than the template fitting approach.

The strengths and weaknesses of the multiband method suggest a hybrid approach to finding periodicity in sparsely-sampled multiband data: a first pass with the fast multiband method, followed by a second pass using the more computationally intensive template-fitting method to select among these candidate periods.
Despite pessimism in previous studies, our experiments with simulated LSST data indicate that such a hybrid approach will successfully identify periods in the majority of RR Lyrae stars brighter than $g\sim 22.5$ in the first months of the survey, and the majority of the faintest detected stars with several years of data.
This finding suggests that the multiband periodogram could have an important role to play in the analysis of variable stars in future multiband surveys.

We have released a Python implementation of the multiband periodogram on GitHub, along with Python code to reproduce all results and figures in this work; this is described in \app{gatspy}.
As we were finalizing this manuscript, we were made aware of a preprint of an independent exploration of a similar approach to multiband light curves \citep{Long2014}; we discuss the similarities and differences between these two approaches in \app{long_comparison}.

{\it Acknowledgments:} JTV is supported by the University of Washington eScience institute, including grants from the Alfred P. Sloan Foundation, the Gordon and Betty Moore Foundation, and the Washington Research Foundation. The authors thank GitHub for providing free academic accounts which were essential in the development of this work.

\bibliographystyle{apj}
\bibliography{paper}

\appendix
\section{Python Implementation of Multiband Periodogram}
\sectlabel{gatspy}
The algorithm outlined in this paper is available in {\tt gatspy}, an open-source Python package for general astronomical time-series analysis\footnote{\url{http://github.com/astroml/gatspy/}} \citep{gatspy}. Along with the periodogram implementation, it also contains code to download all the data used in this work. Code to reproduce this paper, including all figures, is available in a separate repository\footnote{\url{http://github.com/jakevdp/multiband_LS/}}.

{\tt gatspy} is a pure-Python package written to be compatible with both Python 2 and Python 3, and performs fast numerical computation through dependencies on {\tt numpy} \citep{numpy}\footnote{\url{http://www.numpy.org}} and {\tt astroML} \citep{astroML}\footnote{\url{http://www.astroml.org}}, which offer optimized implementations of numerical methods in Python.

The API for the module is largely influenced by that of the {\tt scikit-learn} package \citep{scikit-learn, sklearn_API}\footnote{\url{http://scikit-learn.org}}, in which models are Python class objects which can be fit to data with the \texttt{fit()} method.
Here is a basic example of how you can use {\tt multiband\_LS} to download the data used in this paper, fit a multiband model to the data, and compute the power at a few periods:

\begin{lstlisting}
from gatspy.periodic import LombScargleMultiband
import numpy as np

# Fetch the Sesar 2010 RR Lyrae data
from gatspy.datasets import fetch_rrlyrae
data = fetch_rrlyrae()
t, mag, dmag, filts = data.get_lightcurve(data.ids[0])

# Construct the multiband model
model = LombScargleMultiband(Nterms_base=0, Nterms_band=1)
model.fit(t, mag, dmag, filts)

# Compute power at the following periods
periods = np.linspace(0.2, 1.4, 1000) # periods in days
power = model.periodogram(periods)
\end{lstlisting}

Other models are available as well. For example, here is how you can compute the periodogram under the supersmoother model; this implementation of the supersmoother periodogram makes use of the \texttt{supersmoother} Python package \citep{Vanderplas2015}.

\begin{lstlisting}
from gatspy.periodic import SuperSmoother

# Construct the supersmoother model
model = SuperSmoother()
gband = (filts == 'g')
model.fit(t[gband], mag[gband], dmag[gband])

# Compute power at the given periods
power = model.periodogram(periods)
\end{lstlisting}

The models in the \texttt{gatspy} package contain many more methods, and much more functionality that what is shown here. For updates, more examples, and more information, visit \url{http://github.com/astroml/gatspy/}.

\section{Comparison with \citet{Long2014}}
\sectlabel{long_comparison}
As we were finishing this study, we learned that another group had released a preprint independently addressing the multiband periodogram case, and come up with a solution very similar to the one presented here \citep[][hereafter LCB14]{Long2014}.
They present two methods, the ``Multiband Generalized Lomb-Scargle'' (MGLS) which is effectively identical to the $(1, 0)$ multi-phase model
here, and the ``Penalized Generalized Lomb-Scargle'' (PGLS), which is similar in spirit to our $(0, 1)$ shared-phase model.

In the PGLS model, they start with a multi-phase model, fitting independent $N=1$ term fits to each band, and apply a nonlinear regularization term which penalizes differences in the amplitude and phase. In terms of the formalism used in this work, the PGLS model minimizes a regularized $\chi^2$ of the form
\begin{equation}
  \chi^2_{PGLS} = \sum_{k=1}^K \bigg[~\chi^2_{GLS}(D^{(K)}) + J_A(A^{(k)}) + J_\phi(\phi^{(k)})~\bigg].
\end{equation}
for $K$ bands, where $\chi^2_{GLS}(D^{(K)})$ is the $\chi^2$ of the standard floating mean model on the single-band data $D^{(K)}$, and $J_A$ and $J_\phi$ are regularization/penalty terms which are a function of the amplitude $A^{k}$ and phase $\phi^{(k)}$ of each model. In terms of our linear model parameters $\theta^{(k)}$, this amplitude and phase can be expressed:
\begin{eqnarray}
 A^{(k)} &=& \sqrt{(\theta_1^{(k)})^2 + (\theta_2^{(k)})^2}\nonumber\\
 \phi^{(k)} &=& \arctan(\theta_2 / \theta_1)
\end{eqnarray}
The selected form of these regularization terms penalizes deviations of the amplitude and phase from a common mean between the bands; in this sense the PGLS model can be considered a conceptual mid-point between our shared-phase and multi-phase models.
Within the formalism proposed in the current work, such a mid-point may be alternatively attained by suitably increasing the regularization parameter $\epsilon$ used in our shared-phase model, though the nature of the resulting regularization will differ.

Computationally, the PGLS model requires a nonlinear optimization at each frequency $\omega$, and is thus much more expensive than the straightforward linear optimization of our shared-phase model.
For this reason, LCB14 proposes a clever method by which nested models are used to reduce the number of nonlinear optimizations used: essentially, by showing that the (linear) MGLS $\chi^2$ is a lower-bound of the (non-linear) PGLS $\chi^2$, it is possible to iteratively reduce the number of PGLS computations required to minimize the $\chi^2$ among a grid of frequencies.
Such an optimization could also be applied in the case of our shared-phase model, but is not necessary here due to its already high speed.
Nevertheless, when applying the method to a very large number of light curves, as in e.g.~LSST, such a computational trick may prove very useful.

Given these important distinctions between the models proposed here and in LCB14, in future work we plan to do a detailed comparison of the two means of multiband model regularization.

\end{document}